\documentclass[twocolumn,aps,pre,showpacs,floatfix,groupedaddress]{revtex4-1}
\usepackage{graphics}
\usepackage{graphicx}
\usepackage{amsmath}
\usepackage{amssymb}
\usepackage{ulem}
\usepackage{color}

\begin{document}
%\bibliographystyle{plain}
%\pagestyle{plain}
%\pagenumbering{arabic}%\rmfamily
\title{Interacting Steps With Finite-Range Interactions: Analytical Approximation and Numerical Results}

\author{Diego Felipe Jaramillo}
\email[]{df.jaramillo326@uniandes.edu.co}
\affiliation{Departamento de F\'isica, Universidad de Los Andes, A.A. 4976 Bogot\'a, Colombia}
\author{Gabriel T\'ellez}
\email[]{gtellez@uniandes.edu.co}
\affiliation{Departamento de F\'isica, Universidad de Los Andes, A.A. 4976 Bogot\'a, Colombia}
\author{Diego Luis Gonz\'alez}
\email[]{diego.luis.gonzalez@correounivalle.edu.co}
\affiliation{Departamento de F\'isica, Universidad del Valle, A.A. 25360, Cali, Colombia}
\author{T. L. Einstein}
\email[]{einstein@umd.edu}
\affiliation{Department of Physics and Condensed Matter Theory Center, University of Maryland, College Park, Maryland 20742-4111  USA}

\date{\today}

% Article starts here

\begin{abstract}
We calculate an analytical expression for the terrace-width distribution $P(s)$ for an interacting step system with nearest and next nearest neighbor interactions. Our model is derived by mapping the step system onto a statistically equivalent 1D system of classical particles. The validity of the model is tested with several numerical simulations and experimental results. We explore the effect of the range of interactions $q$ on the functional form of the terrace-width distribution and pair correlation functions. For physically plausible interactions, we find modest changes when next-nearest neighbor interactions are included and generally negligible changes when more distant interactions are allowed.  We discuss methods for extracting from simulated experimental data the characteristic scale-setting terms in assumed potential forms.

\end{abstract}
%\pacs{89.75.Kd,68.55.A-,05.40.-a,81.16.Rf}
%89.75.Kd    Pattern formation in complex systems
%68.55.Ac    Nucleation and growth: microscopic aspects
%05.40.-a    Fluctuation phenomena, random processes, noise, and Brownian motion
%81.16.Rf    Nanoscale pattern formation
%81.15.Aa    Theory and models of film growth
%05.10.Gg    Stochastic analysis methods (Fokker-Planck, Langevin, etc.)
%05.45.Mt    Quantum chaos; semiclassical methods
%81.10.Aj    Crystal growth, theory and models of
%05.70.Ln    Irreversible thermodynamics
%68.35.-p    Solid surfaces and solid-solid interfaces: Structure and energetics
%68.43.Jk    Diffusion of adsorbates, kinetics of coarsening and aggregation
%82.20.Pm    Rate constants, reaction cross sections, and activation energies

%{\bf Keywords:} Interacting steps, terrace width distribution, terrace-step-kink model.\\

\maketitle
% Article starts here

%\begin{abstract}
%\end{abstract}

%{\bf Keywords:} Systems out of equilibrium, random matrices, Wigner surmise.\\
\section{Introduction}
The equilibrium properties of steps on surfaces have been the subject of study for at least half a century \cite{FM63}.
During the last two decades, interest in steps has burgeoned \cite{jaysaam84,akutsu1,akutsu2,einstein1,einsteinpoin,williams1,einstein2,einstein5,einstein6,einstein8},
principally because of the important role they play in epitaxial growth, surface transport, catalysis, etc.  Those properties have great importance in the construction of nano and micro-electronic devices.

Steps can be created during sample preparation, such as by cutting a material at a miscut angle with respect to a closely packed plane. In the case of molecular-beam epitaxy (MBE), the steps act as sinks because they are the most favorable place for attachment, allowing some control of the morphology of the surface during growth and creating specified uniaxial defects rather than random growth of domains that progressively degrade uniformity.

Advances in experimental techniques have allowed the quantitative measurement of some statistical properties of these stepped surfaces \cite{williams1}. One of the most important equilibrium statistical properties is the terrace-width distribution (TWD), $P(s)$. Here
$s=S/\langle S\rangle$ is the scaled width, with $\left\langle S\right\rangle$ the average of $S$. The TWD has relevant information about the interaction potential between steps \cite {williams2,margret,einstein15,einstein16}; the TWD narrows as step-step repulsions increase.  %and about the mass transport mechanisms on terraces

Recent connections between theory and experiments have often relied on the generalized Wigner surmise (GWS) \cite{einstein11,einstein17}
\begin{equation}\label{gws}
P_\varrho(s)=a_\varrho s^\varrho {\rm e}^{-b_\varrho s^2},
\end{equation}
where %$b_\varrho =\left[\frac{\Gamma({\varrho+2\over2})}{\Gamma({\varrho+1\over2})}\right]^2$
$b_\varrho =\left[\Gamma({\varrho+2\over2})/\Gamma({\varrho+1\over2})\right]^2$
and %$ a_\varrho\! =\! \frac{2b_\varrho^{(\varrho+1)/2}}{\Gamma({\varrho+1\over2})}$
$ a_\varrho\! =\! 2b_\varrho^{(\varrho+1)/2}/\Gamma({\varrho+1\over2})$
are normalization constants which ensure $\left\langle s^n \right\rangle$=1 for $n=0,1$.

In particular, for the typical case where stress dipoles at steps lead them to interact through a potential $V(S)=A/S^{2}$, it has been shown that there is a connection between $A$ and the exponent $\varrho$ of the GWS for the special cases $\varrho=1$, $2$ and $4$:  Explicitly we have that the dimensionless repulsion strength
\begin{equation}\label{eca}
\tilde{A}\equiv\tilde{\beta}\,A\,\beta^2=\frac{\varrho}{2}\left(\frac{\varrho}{2}-1\right),
\end{equation}

\noindent where $\tilde{\beta}$ is the step stiffness, and $\beta=1/{k_B}T$ is the inverse thermal energy. The remarkable connection provided by Eq.~(\ref{eca}) can be found by mapping the interacting steps system onto the Calogero-Sutherland model \cite{Calogero,Sutherland}, in which the particles interact through a potential $\tilde{A}/S^2$. It is clear that the case $\varrho=2$ corresponds to non-interacting steps ($A=0$), $\varrho<2$ to attractive steps ($A<0$) and $\varrho>2$ to repulsive steps ($A>0$). The Calogero-Sutherland model can be solved analytically (the Hamiltonian is integrable) for the special cases of $\tilde{A}=-1/4,0,2$, and Eq.~(\ref{gws}) provides an excellent approximation to the spacings between adjacent particles \cite{einstein1,einsteinpoin,haake}. While the justification of Eq.~(\ref{eca}) is not so firmly established for arbitrary values of $\tilde{A}$, the GWS has nonetheless proved to be an excellent tool to describe theory, experimental and numerical results \cite{einstein12,einstein13,einstein9,einstein10,hailu,einstein20}.

One of the simplest models to describe fluctuations on steps is the terrace-step-kink model (TSK) \cite{kossel,stranski,jaysaam84,einstein6, einstein5}. In this model, the only excitations taken into account are the kinks along the steps. In particular, vacancies and adatoms on terraces are neglected. These simplifications restrict the applicability of the TSK model to low temperatures (relative to the roughening temperature of the terraces). There are more sophisticated approaches \cite{karim,szalma,ajmi} based on kinetic Monte Carlo simulations \cite{voter} of solid-on-solid (SOS) models \cite{gilmer} which take into account more thermal excitations, but they require more computational resources. Moreover, the simplicity of the TSK results in better statistics for $P(s)$ than the SOS models in the low-temperature limit.

Our main objective is to calculate the TWD for arbitrary interaction potentials in the case of finite-range interactions, i.e., when each step interacts with a number $2\,q$ of its neighbors through a potential $V(S)$. In addition to the intrinsic interest of this calculation, the question is very important for doing Monte Carlo simulations to test predictions of models like Calogero-Sutherland, which assume that all steps interact.  When step-step interactions are included in simulations, invariably only nearest-neighbor interactions are included \cite{hailu,einstein20,weeks,uwaha03,frisch05}.  Including second or third neighbor interactions in the simulations algorithm would be cumbersome but feasible.  However, doing so invites questions of whether such longer-range interactions are screened by intervening steps or have their ``bare" form.

Another important issue is what happens if the step interactions are not of the generic inverse-square form, in particular if they decay more slowly, often leading to instabilities.  In accounting for the idiosyncratic step network on Au(110) and Pt(110), Carlon and van Beijeren find what amounts to an $S^{-1}$ repulsion \cite{vanBeij96}.  Stress domains lead to logarithmic interactions, notably in the case of terraces with alternating mutually-perpendicular domains on vicinal Si(100) surfaces \cite{alerhand88,alerhand90}.

To proceed, we map the step system onto a 1D classical system of interacting particles. One advantage of this approach is that it is always possible to find analytically (in Laplace space) the spacing-distribution functions for these 1D classical systems. In particular, the nearest-neighbor distribution (TWD) is easy to obtain. Additionally, it allows one to determine how relevant are the interactions beyond the nearest-neighbors in the functional form of the TWD. The applicability of our model is tested with several Monte Carlo simulations and some experimental results. This paper is organized as follows. In Sec. II we describe the terrace-step-kink (TSK) model for interacting steps, in Sec. III we develop an analytical model for the TWD for the cases $q=1$ and $q=2$. In this section, the case of arbitrary values of $q$ is also discussed. Finally in Sec. IV we provide some conclusions.

\section{Terrace-step-kink model}
Since overhangs are prohibited in the TSK model, the position of the $i$-th step can be described by a function $x_i(y_n)$ where we have used ``Maryland notation'' \cite{einstein1}, in which $\hat{y}$ is the ``time-like'' direction along the step. Then, the indices satisfy $i\in[1,N]$, with $N$ the number of steps, and $n\in[1,L_y]$  with $L_y$ the length of the lattice in $y$ direction. The fermionic non-touching condition imposes the additional restriction $x_i(y_n)<x_{i+1}(y_n)$ for all $i$ and $n$. For the sake of simplicity, in our model the lattice constant is set to unity.  We also impose periodic boundary conditions in both $x$ and $y$ directions.

In the TSK model, the Hamiltonian of a system of interacting steps can be written as \cite{rajesh}
\begin{equation}\label{ham}
H=\sum_{y_n=1}^{Ly}\left(\sum_{i=1}^{N}\epsilon_k\,\xi_{i}+\sum_{i=1}^N\sum_{j=i+1}^{i+q} V(\zeta_{i,j})\right)
\end{equation}
where $\xi_{i}(y_{n+1},y_{n})=\left|x_i(y_{n+1})-x_i(y_{n})\right|$, $\zeta_{i,j}(y_n)=\left|x_i(y_{n})-x_j(y_{n})\right|$, $\epsilon_k$ is the energy required to form a unit-length kink, and $V(\zeta)$ is the interaction potential between steps. As mentioned previously, $q$ is the range of interaction. For $q=1$, we have nearest-neighbor interactions while for $q=(N-1)/2$ each step interacts with all its neighbors (full-range interactions).

The most studied case corresponds to $V(\zeta)=0$, which is usually called non-interacting steps. However, we emphasize that even in this case the steps represented by Eq.~(\ref{ham}) interact entropically due the non-touching condition, taking the well-defined thermodynamic form when
\begin{equation}
y_{coll}=\left(\frac{\left\langle S\right\rangle}{2\,b(T)}\right)^2<L_y\,\,\,\,\mathrm{with}\,\,\,\,b^2(T)=\frac{2}{2+e^{\beta\,\epsilon_k}}.
\end{equation}
For $L_y$ smaller than $y_{coll}$ the steps fluctuate independently of each other \cite{bustingorry}. The case of non-interacting steps is well described by the free-fermion analogy. In this picture, the steps are modeled as world lines of free spinless fermions \cite{einstein6}. This analogy leads to the use of the Wigner surmise with $\varrho=2$ to describe $P(s)$ \cite{einstein6}.

Dyson showed \cite{dyson1,dyson2} long ago that the statistical behavior of a 1D free-fermion system (non-interacting steps) is equivalent to that of a 1D system of classical Brownian particles \cite{mehta}. Then, in terms of a 1D classical system, the non-interacting steps can be interpreted as the world lines of a system with $N$ particles on a ring which interact via a logarithmic potential at an inverse temperature $\beta=2$. Explicitly, the Hamiltonian of this system is given by
\begin{equation}\label{hpotlog}
H=-\frac{1}{2}\sum^{N}_{i=1}\sum^{N}_{j=1}\mathrm{ln}\left|z_i(t)-z_j(t)\right|,
\end{equation}
where $z_i(t)$ is the position of the $i$-th particle at time $t$. If we interpret the time axis as the $y$-axis of the vicinal surface, we can represent the steps in the TSK model as the time evolution of the positions of particles in a 1D classical system (after making the step-continuum approximation or working in discrete time). In this equivalent system the interparticle gap size distribution plays the role of the TWD.

\section{Analytical model}\label{secPV}
To map the interacting step system onto a classical 1D system of interacting particles, we consider $N$ particles can move around a circle with circumference $L_x=\left\langle S\right\rangle N$, with $\left\langle S\right\rangle$ the average distance between particles. Periodic boundary conditions are perforce imposed, that is, $z_{N+j}=z_j$, where $z_j$ is the position of the $j$-th particle. The system is in equilibrium at an inverse temperature $\beta$. Below we use the formalism proposed in Ref.~\cite{Bogomolny} to calculate the interparticle gap size distribution  $P(s)$ for different ranges of interaction $q$.

\subsection{Nearest-neighbor interactions $q=1$}
Now we consider the simplest case where the particles/steps interact with their nearest-neighbors through of an arbitrary potential $\tilde{V}(S;{\cal A})\equiv V(S)$, where ${\cal A}$ is a dimensionless parameter which determines the strength of the interaction in such way that ${\cal A}=0$ implies non-interacting steps, viz.  $A=0$. In general,  the interaction potential between steps $V(S;A)$ and the one for the classical particle system, $\tilde{V}(S;{\cal A})$, are related according to
\begin{equation}\label{ren}
\tilde{V}(S;{\cal A})=f({\cal A})V(S;A),
\end{equation}
where both ${\cal A}=g(A)$ and $f({\cal A})$ are unknown functions. This means that there is a scale relation between the two potentials, i.e. the functional form of the interaction potential is the same in both cases.

To map the step system onto a 1D classical interacting particle system, we use the Hamiltonian

\begin{equation}\label{hp1}
H=-\frac{1}{2\,\beta}\sum^{N}_{i=1}\sum^{N}_{j=1}\mathrm{ln}\left|\zeta^2_{i,j}\right|+\sum^{N}_{i=1}\tilde{V}(\zeta_{i,i+1};{\cal A}),
\end{equation}
where $\zeta_{i,j}\equiv\zeta_{i,j}(t)=z_i(t)-z_j(t)$. The first term in Eq.~(\ref{hp1}) models the entropic repulsion between steps while the second one takes into account the energetic interaction between steps. We have full-range interactions for the logarithmic potential but for $\tilde{V}(\zeta;{\cal A})$ we just have nearest-neighbor interactions. Instead of this potentially difficult scenario and following Ref.~\cite{gonzalez} (cf.\ Eq.~(25) therein), we propose the effective Hamiltonian
\begin{equation}\label{hpeff1}
H_{\mathrm{eff}}=\frac{1}{\beta}\sum^{N}_{i=1}\left[K S^2_i-\mathrm{ln}\left(S^2_i\right)\right]+\sum^{N}_{i=1}\tilde{V}(S_i;{\cal A}),
\end{equation}
where $S_i=\zeta_{i,i+1}$ and $K \equiv K({\cal A})$ is a function of ${\cal A}$ which is determined by the normalization conditions as in Eq.~(\ref{gws}). The advantage of Eq.~(\ref{hpeff1}) over Eq.~(\ref{hp1}) lies in the fact that both potentials have the same range of interaction ($q=1$), allowing an easier computation of $P(s)$. According to Ref.~\cite{gonzalez}, we can expect that the TWD given by Eq.~(\ref{hpeff1}) for ${\cal A}=0$ reduces to the GWS with $\varrho=2$, as required.

As shown in Appendix \ref{app3}, the TWD for the system described by Eq. (\ref{hpeff1}) is given by
\begin{equation}\label{TWDp1}
P(s)=\frac{1}{\tilde{f}(c)}s^2e^{-\Gamma\,s^2-\beta\,v(s;{\cal A})-c\,s}
\end{equation}
where $c$ and $\tilde{f}(c)$ are given by the normalization conditions and $\upsilon(s;{\cal A})$ is the step-step interaction potential in dimensionless form. Consequently, Eq.~(\ref{TWDp1}) has just the one free parameter ${\cal A}$.

From now on, we will consider interaction potentials which satisfy $\tilde{V}(S;{\cal A})\rightarrow0$ for $S\rightarrow\infty$. Consequently, in this limit the TWD behaves as
\begin{equation}
P(s)\approx\frac{1}{\tilde{f}(c)}e^{-\Gamma\,s^2}.
\end{equation}
This means that $v(s;{\cal A})$ does not change the functional form of the TWD for large values of $s$ compared with the case of non-interacting steps ($A=0$), which, in turn, is well described by Eq.~(\ref{gws}) with $\varrho=2$. This is not an unexpected result because $v(s;{\cal A})$ decays as $s\rightarrow \infty$. Thus, we can expect that the interaction potential only has significant effects on the TWD for small and intermediate values of $s$, depending on how fast the interaction potential goes to zero as $s$ increases.

For small values of $s$, the functional form of the TWD depends strongly on $v(s;{\cal A})$. For example consider the general interaction potential $v(s;{\cal A})={\cal A}\,s^{-\gamma}$. For $s\ll1$, this particular potential leads to
\begin{equation}
P(s)\approx\frac{1}{\tilde{f}(c)}e^{-\beta\,{\cal A}\,s^{-\gamma}},
\end{equation}
which is clearly different from the behavior $P(s)\approx a_{\varrho}s^{\varrho}$ predicted by the GWS. However, for sufficiently large values of $\gamma$ the effect of the $v(s;{\cal A})$ becomes important just for small values of $s$. This justifies the use of the GWS to fit the TWD for rapidly decreasing potentials and explains why it gives excellent results for $v(s;{\cal A})={\cal A}\,s^{-\gamma}$ with $\gamma=2$ and (unphysically) 3; see, for example, Ref.~\cite{einstein17}.

\begin{figure}[htp]
\begin{center}
\includegraphics[scale=0.28]{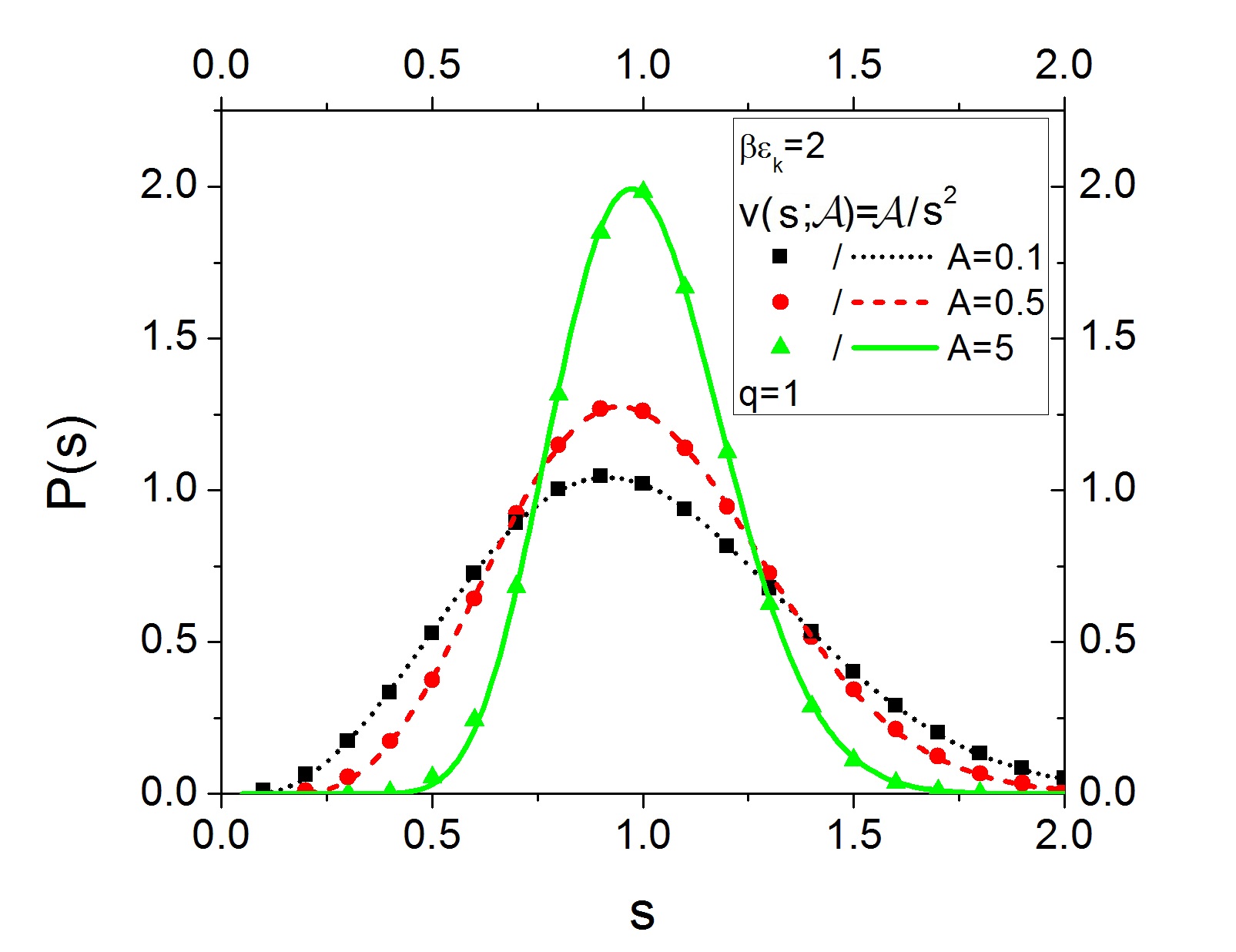}\\
\caption{(Color online) TWD for different values of $A$ with $v(s;{\cal A})={\cal A}\,s^{-2}$ and nearest-neighbor interactions $q=1$ . In all figures we include  the values of $A$ used in the numerical simulation instead of the ones of ${\cal A}$.
The relation between both constants is given implicitly by Eq.~(\ref{ren}). }
\label{pfita}
\end{center}
\end{figure}

As mentioned before, Eq.~(\ref{TWDp1}) has just one free parameter. Nonetheless, it describes quite well the data obtained from the numerical simulation of the TSK model for different potentials. The case of $v(s;{\cal A})={\cal A}\,s^{-2}$ is shown in Fig.~\ref{pfita} for different values of ${\cal A}$. The agreement between numerical and analytical results given by Eq.~(\ref{TWDp1}) is excellent \cite{simu}.

\begin{figure}[htp]
\begin{center}
\includegraphics[scale=0.28]{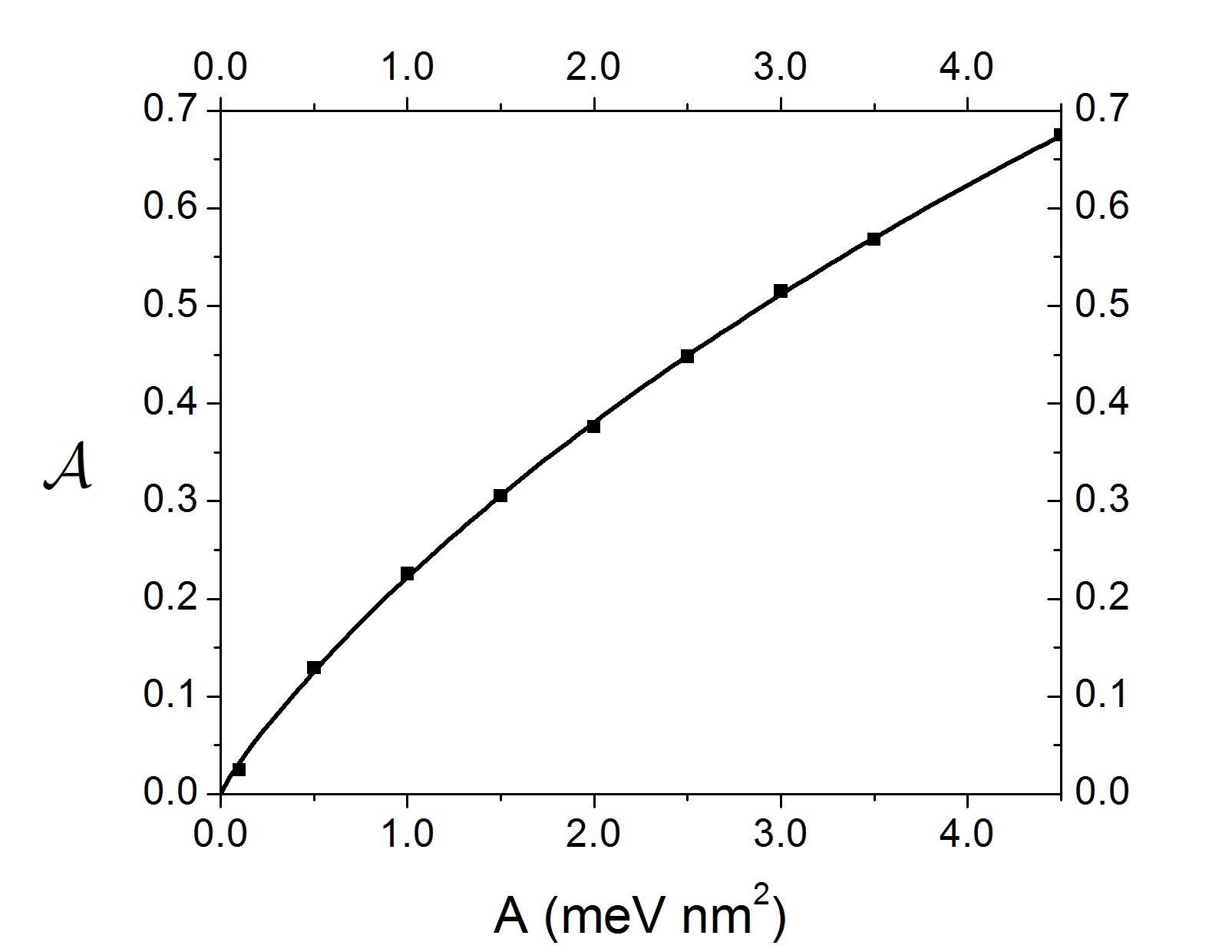}\\
\caption{Relation between ${\cal A}$ and $A$ for the potential $v(s;{\cal A})={\cal A}\,s^{-2}$ and nearest-neighbor interactions $q=1$. The continuum line is given by ${\cal A}^{\frac{1}{\eta}}=\nu\,\mathrm{ln}(\chi A+1)$ with $\nu\approx 1.05$, $\eta \approx 0.87$ and $\chi \approx 0.19 (meV nm^{2}) ^{-1}$ while the dots represent the numerical data.}
\label{amvsa}
\end{center}
\end{figure}

In general, the function ${\cal A}=g(A)$ cannot be determined easily from analytical calculations. However, we find that the empirical relation ${\cal A}^{\frac{1}{\eta}}=\nu\,\mathrm{ln}(\chi A+1)$ fits well the numerical relation between ${\cal A}$ and $A$ found from the numerical data, as shown in Fig. \ref{amvsa}.

Now, we consider a more general interaction between steps. We adopt the potential
\begin{equation}\label{vab}
v(s;{\cal A},B)=\frac{{\cal A}}{s^2}+\frac{B\,\cos(\omega\,s+\phi)}{s^{3/2}},
\end{equation}
which is characteristic of vicinal surfaces with both elastic repulsion and surface-state mediated electronic interactions \cite{redfeld,einstein30,einstein17,einstein3,einstein4,einstein14}. In this expression, ${\cal A}$ and $B$ are determined by the elastic repulsion and by the coupling to the surface state, respectively; $\omega$ is proportional to the Fermi wavevector; and $\phi$ is a phase shift, for more information see Ref.~\cite{redfeld,einstein30,einstein17}.

\begin{figure}[htp]
\begin{center}
\includegraphics[scale=0.28]{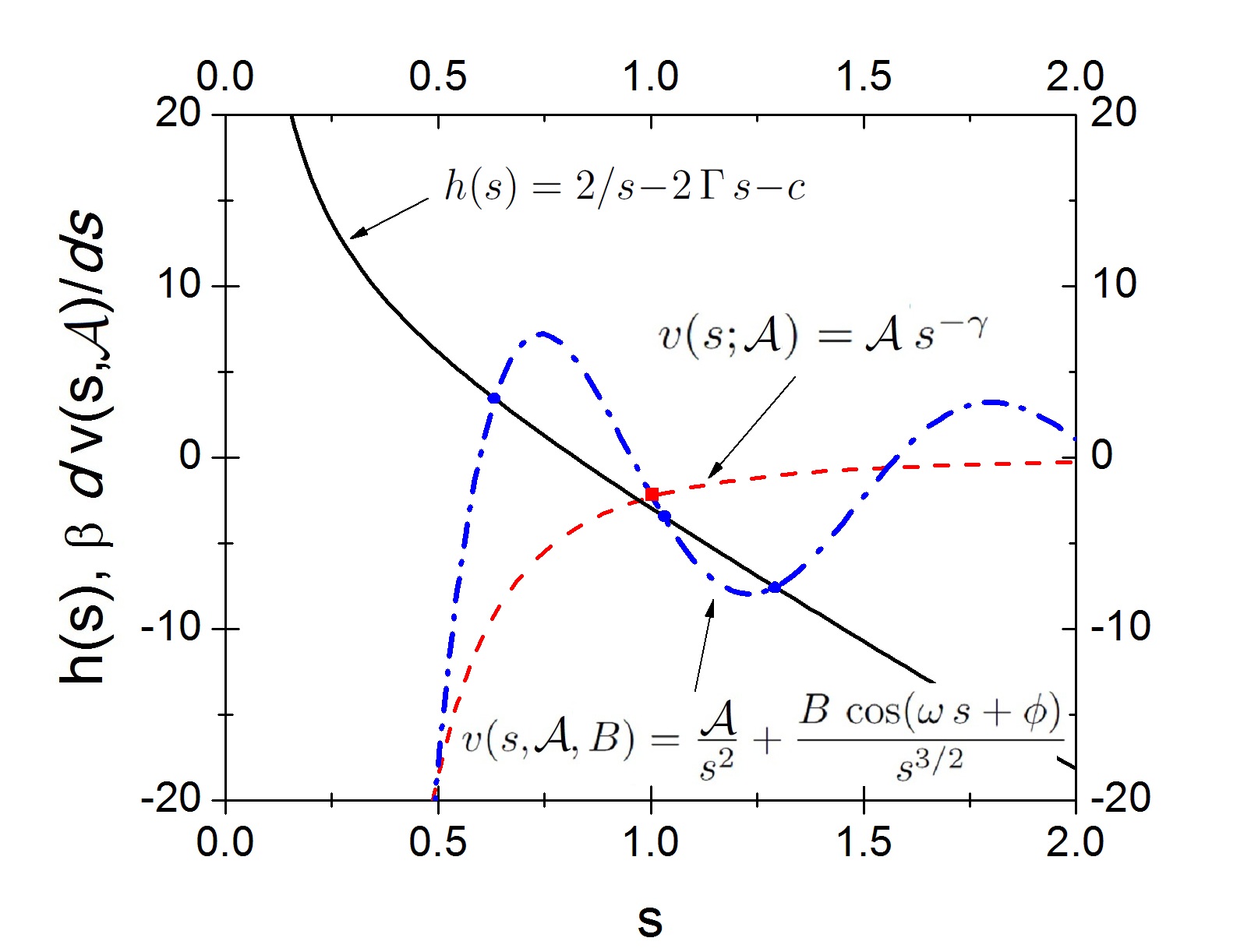}\\
\caption{(Color online) The interaction given by Eq.~(\ref{vab}) may generate more than one maximum in the TWD.}
\label{root}
\end{center}
\end{figure}

\begin{figure}[htp]
\begin{center}
\includegraphics[scale=0.28]{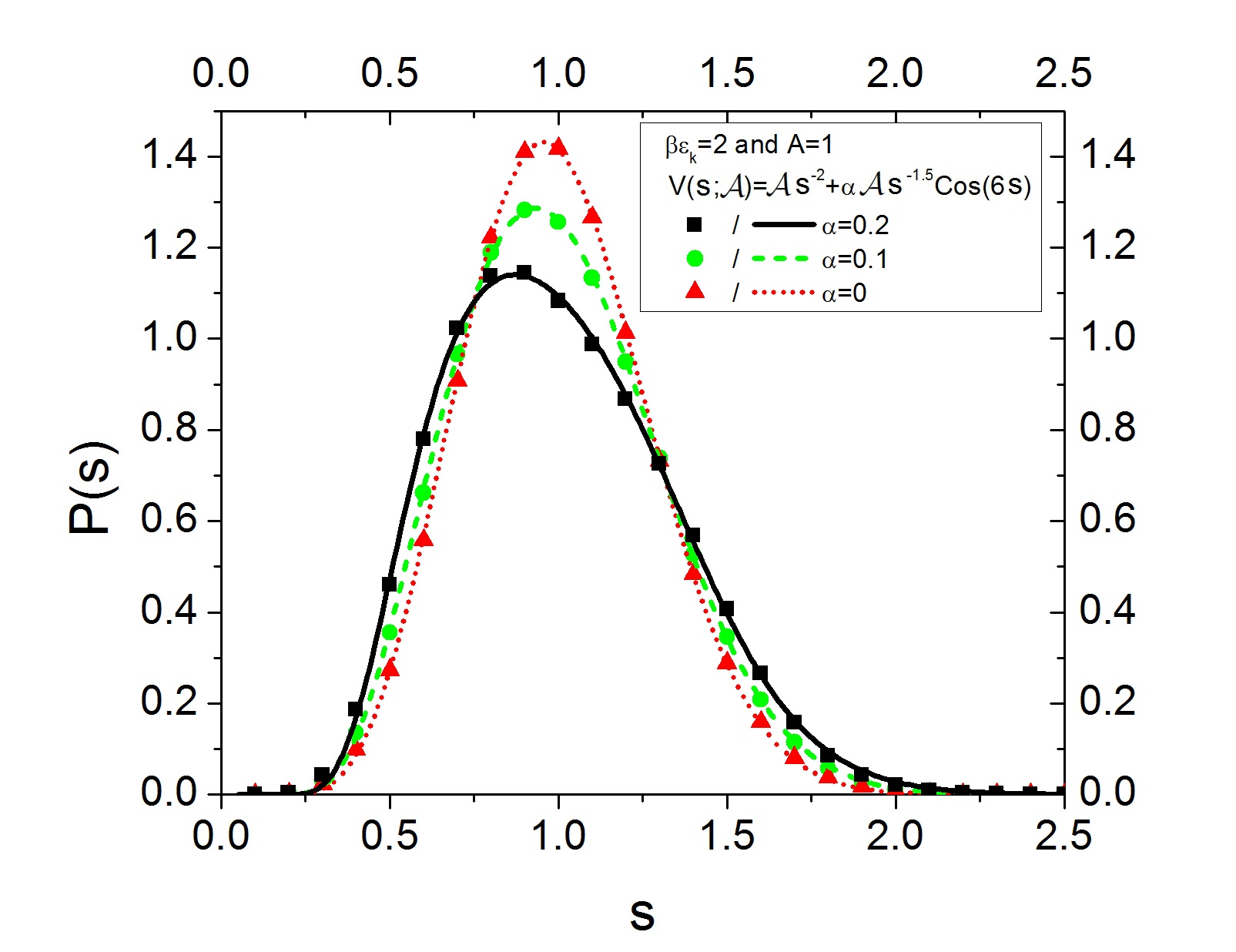}\\
\caption{(Color online) TWD for different values of $B=\alpha\,{\cal A}$ with $v(s;{\cal A},B)=\frac{{\cal A}}{s^2}+\frac{B\,\cos(\omega\,s+\phi)}{s^{3/2}}$ and nearest-neighbor interactions $q=1$.}
\label{pfitb}
\end{center}
\end{figure}

One interesting feature of this potential is the appearance of more than one maximum in the TWD. This scenario appears, for example, in kinetic Monte Carlo (KMC) simulations of solid-on-solid models where step bunching is present \cite{pimpi} as well as in experiments. From Eq.~(\ref{TWDp1}), it is clear that the critical points of $P(s)$ are given by
\begin{equation}
\frac{2}{s}-2\,\Gamma\,s-c-\beta\frac{dv(s;{\cal A})}{ds}=0.
\end{equation}
The function $h(s)=2/s-2\,\Gamma\,s-c$ decreases monotonically, because $\Gamma>0$. In the particular case of $v(s;{\cal A})={\cal A}\,s^{-\gamma}$, the function $\beta\frac{dv(s;{\cal A})}{ds}$ increases monotonically, allowing just one maximum in the TWD. However, for the potential given by Eq.~(\ref{vab}), $\beta\frac{dv(s;{\cal A})}{ds}$ exhibits oscillatory behavior, which can lead to more than one maximum in the TWD. A sketch of this discussion is shown in Fig.~\ref{root}, where there are three critical points (two maxima and one minimum); for the potential $v(s;{\cal A})={\cal A}\,s^{-\gamma}$, there is just one critical point (maximum).  As shown in Fig.~\ref{pfitb}, Eq.~(\ref{TWDp1}) also gives excellent results for this potential.

\begin{figure*}[htp]
\begin{center}
$\begin{array}{cc}
\includegraphics[scale=0.28]{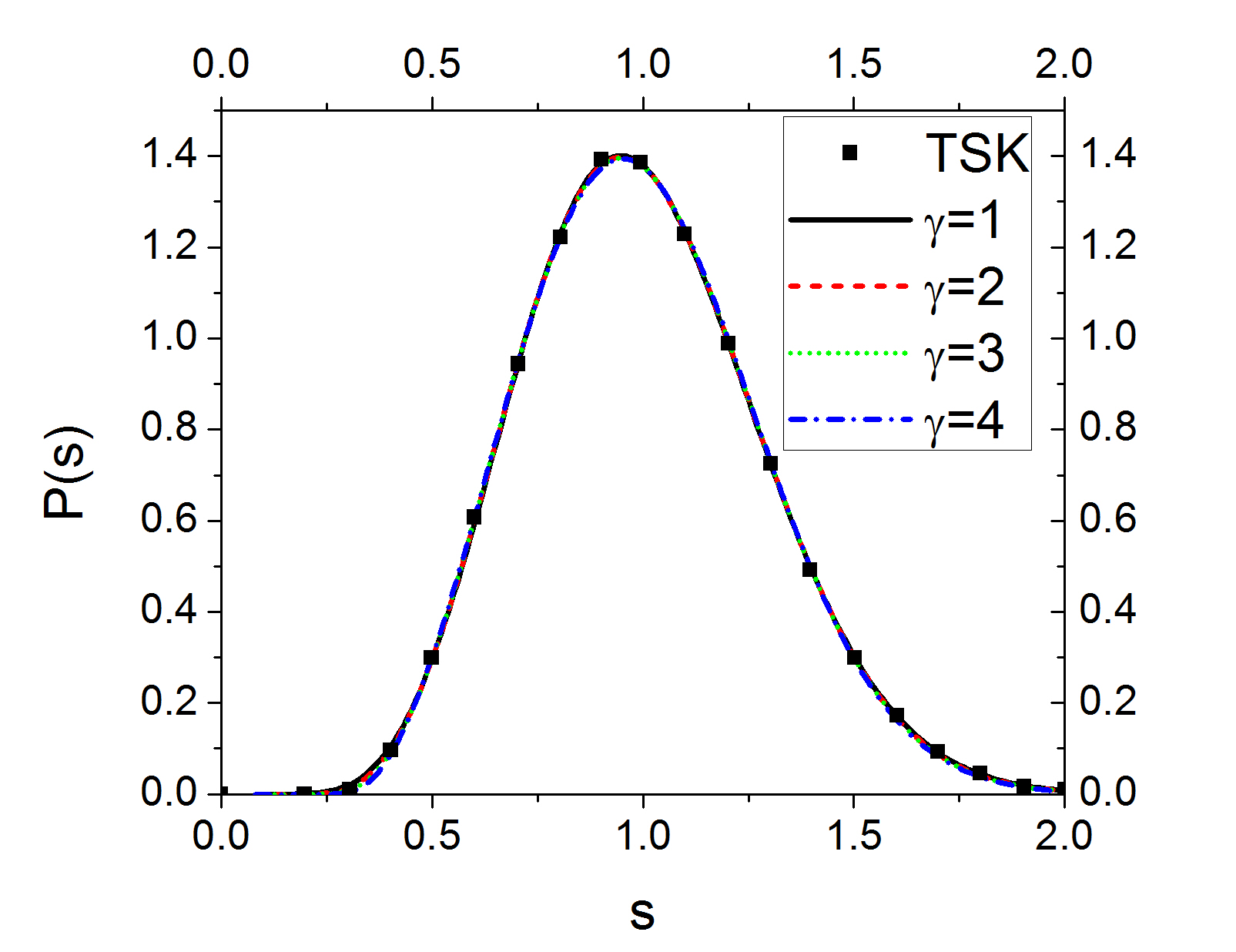} &
\includegraphics[scale=0.28]{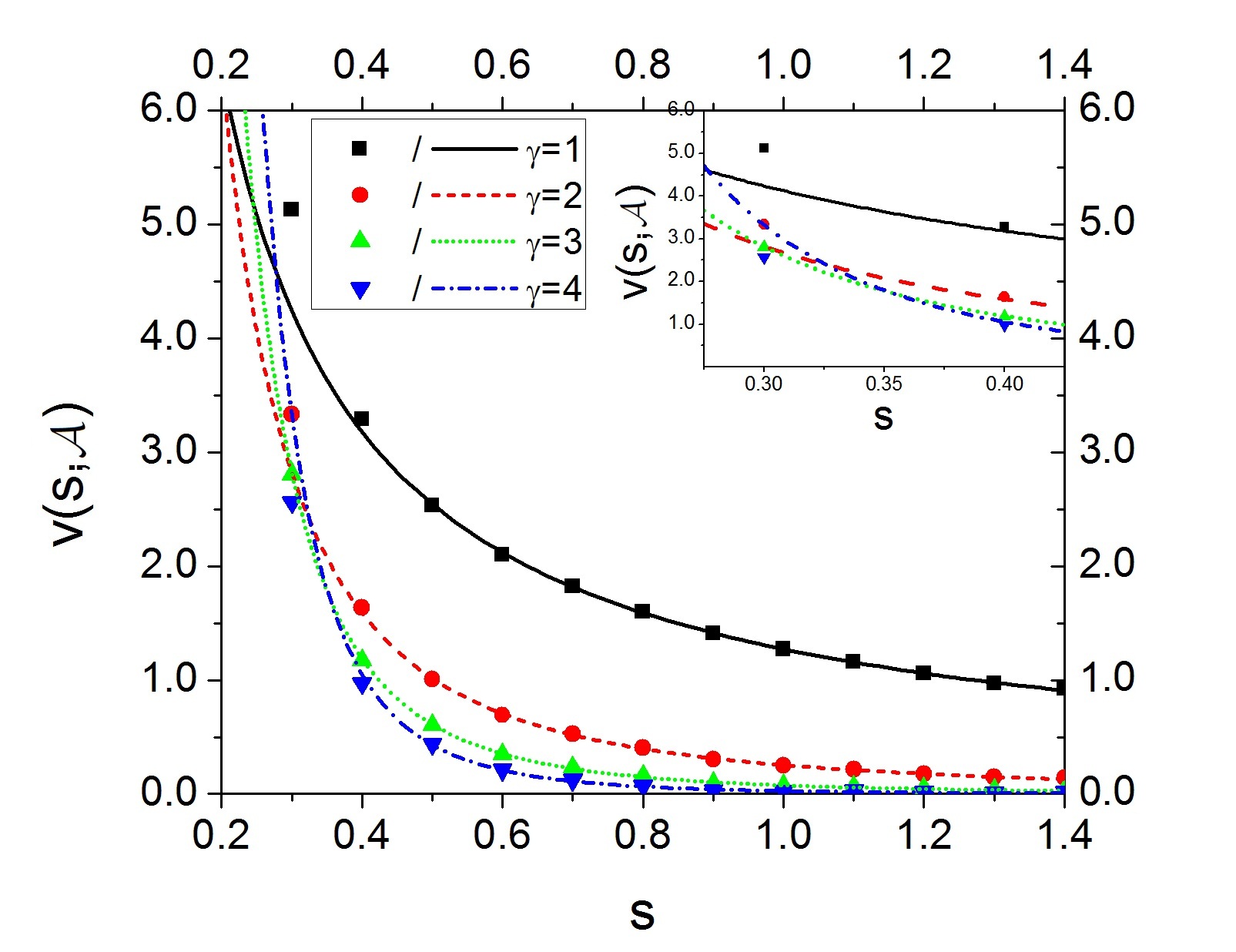}\\
(a) & (b) \\
\end{array}$
\end{center}
\caption{(Color online) Determination of the interaction potential from numerical data from Fig. 2 of Ref. \cite{einstein17} for $V(S) \propto S^{-3}$. (a) All values of $\gamma$ describe the TWD well. (b)  The interaction potential for large values of $s$ is well described in all cases. However, just $\gamma=3$ gives the appropriate fits for the numerical data for $P(s)$ and $v(s;{\cal A})$ for the entire range of $s$.}
\label{s3}
\end{figure*}

Another important advantage of this formalism is that, from Eq.~(\ref{TWDp1}), it is possible to determine the interaction potential $v(s;{\cal A})$ from numerical or experimental data for the TWD. In fact, Eq~(\ref{TWDp1}) can be written in the form
\begin{equation}\label{exppot}
\beta\,v(s;{\cal A})=-\mathrm{ln}\left(\frac{\tilde{f}(c)\,P(s)}{s^2\,e^{-\Gamma\,s^2}}\right)-c\,s,
\end{equation}
By using Eq.~(\ref{exppot}) it is possible to extract $v(s;{\cal A})$ directly from $P(s)$. Nevertheless, as was pointed out in Ref.~\cite{einstein17} this is not a trivial matter even if good quality data are available. As a first example we consider the numerical data given in Fig. 2 of Ref.~\cite{einstein17} for the potential $V(S)\propto S^{-3}$. These data are represented by small squares in Fig.~\ref{s3}. In order to calculate $v(s;{\cal A})$, we proceed as follows. First, we assume a functional form for the interaction potential. For this particular example, we use $v(s;{\cal A})={\cal A}\,s^{-\gamma}$ where ${\cal A}$ and $\gamma$ are parameters to be determined. Second, we select a value of $\gamma$ and then we perform the fit of Eq.~(\ref{TWDp1}) to the data. At the end of this step, we have the parameters ${\cal A}$ and $\gamma$, which define the pre-established form of the interaction potential. The third and final step is to calculate $v(s;{\cal A})$ from Eq.~(\ref{exppot}) in order to check consistency with the pre-established form $v(s;{\cal A})$ of the potential. In Fig.~\ref{s3} the results of fits for $\gamma=1$ to 4 are shown. All of them describe the TWD well; in fact the lines are almost indistinguishable except in the region $s\le 0.3$. However, as we can see in the inset of Fig.~\ref{s3} b), the results for $v(s;{\cal A})$ are consistent for small values of $s$ only in the case $\gamma=3$ \footnote{The case $\gamma=3$ has a limited physical interest but it was included to illustrate the applicability of the model.}. In Ref.~\cite{einstein17} a different approach was used to calculate the same potential leading to the erroneous result $\gamma=2$, while the formalism presented here gives the correct value $\gamma=3$.

\begin{figure}[htp]
\begin{center}
\includegraphics[scale=0.28]{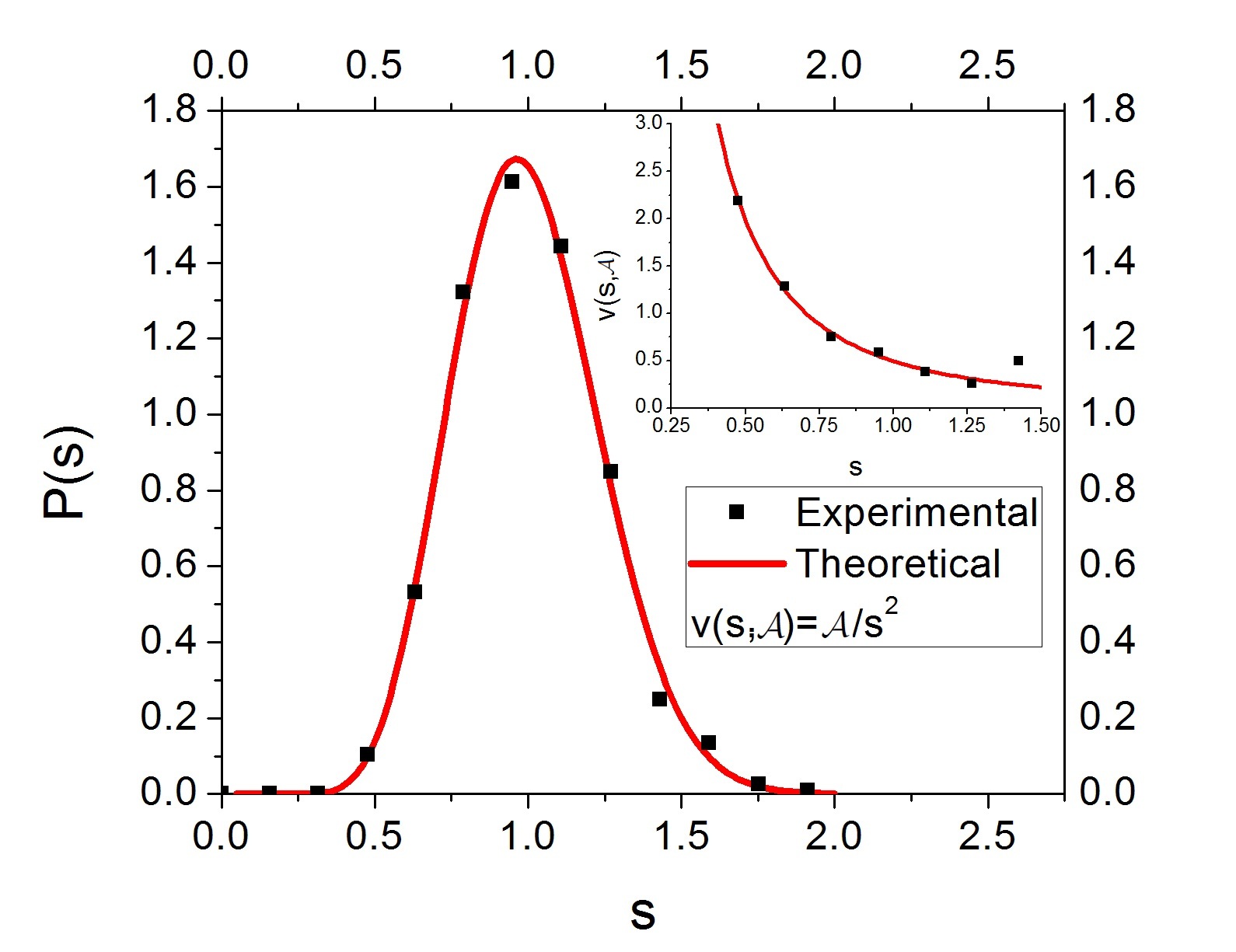}\\
\caption{(Color online) Determination of the interaction potential from experimental data. Excellent agreement is found with $v(s;{\cal A})\propto s^{-2}$.}
\label{wang}
\end{center}
\end{figure}

As an additional example we calculate the potential from the data reported in Fig. 2 of Ref.~\cite{einstein16}. As shown in Fig.~\ref{wang}, both $P(s)$ and $v(s;{\cal A})$, are well described assuming $v(s;{\cal A})\propto s^{-2}$.

Sometimes the results of the experiments are given through the pair correlation function $g(r)$ instead of the TWD \cite{einstein15}. By definition $g(r)$ is the probability of finding another step a specified distance away, regardless of how many steps might lie between them. In the case of a one-dimensional free fermion system, i.e., for non-interacting steps ($A=0$), $g(r)$ can be calculated easily \cite{einstein6,Nijs,Kogut,de Gennes}
\begin{equation}\label{grff}
g(r)=1-\left(\frac{\sin(\pi\,r)}{\pi\,r}\right)^2.
\end{equation}

Unfortunately, the case of interacting steps $v(s;{\cal A})\ne0$ provides a more difficult scenario where $g(r)$ cannot be calculated explicitly in tractable form \cite{forrester,ha}. However, the formalism used to describe the TWD can be extended to find an approximation for $g(r)$: let $p^{(n)}(s)$ be the probability density that the normalized distance between two steps is $s$ under the condition that between them there are $n$ additional steps, see Appendix A. This immediately implies that the TWD is given by $P(s)\equiv p^{(0)}(s)$. Additionally, from $p^{(n)}(s)$ it is possible to recover $g(r)$ through
\begin{equation}\label{greq}
g(r)=\sum_{n=0}^{\infty}p^{(n)}(s).
\end{equation}
We can expect that a good approximation for $p^{(n)}(s)$ gives a good description of $g(r)$. As usual the case $A=0$ is the easiest.

\begin{figure*}[htp]
\begin{center}
$\begin{array}{cc}
\includegraphics[scale=0.26]{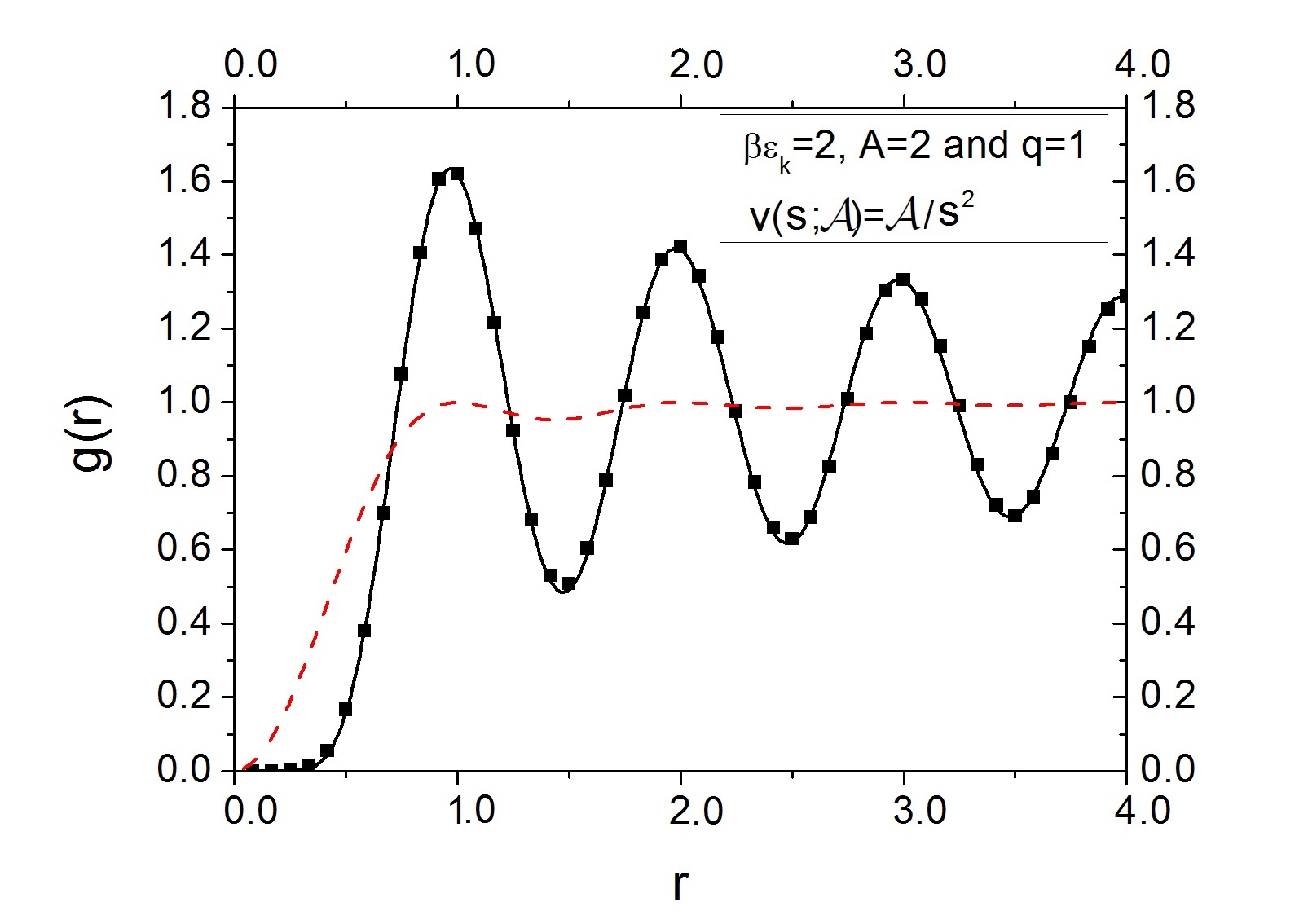} &
\includegraphics[scale=0.26]{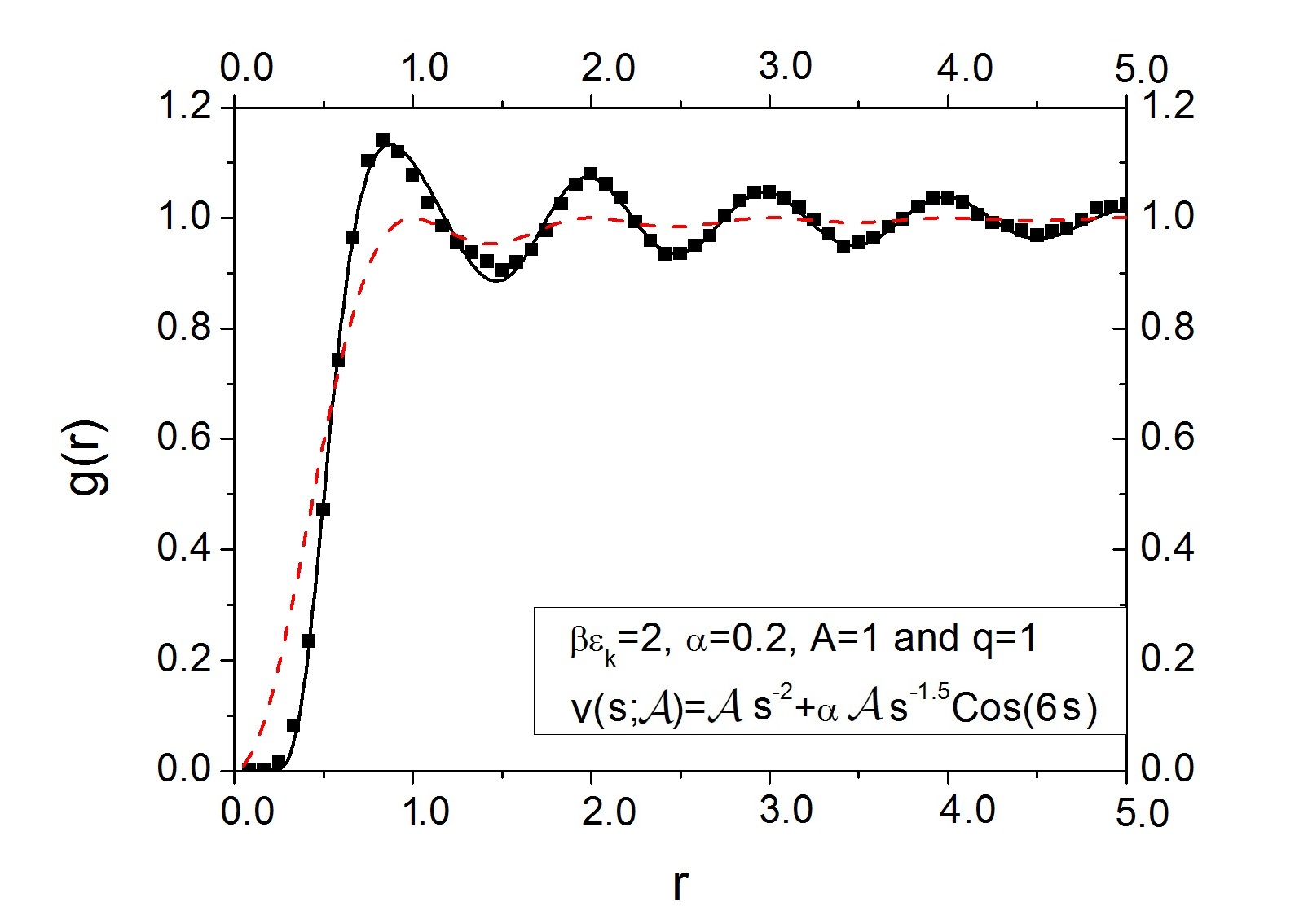}\\
(a) & (b) \\
\end{array}$
\end{center}
\caption{(Color online) Pair correlation function $g(r)$ for interacting steps with a) $v(s;{\cal A})={\cal A}/s^2$ and b) $v(s;{\cal A})$ given by Eq. ~(\ref{vab}). The dashed line corresponds to Eq.~(\ref{grff}) which gives $g(r)$ for ${\cal A}=0$.}
\label{gr}
\end{figure*}

Abul-Magd \cite{abul} showed that, for a one-dimensional free fermion system, $p^{(n)}(s)$ can be approximated by Eq.~(\ref{gws}) taking
\begin{equation}\label{rhon}
\varrho_n=n^2+4\,n+2.
\end{equation}
This approximation assumes that the $p^{(n)}(s)$ for any $n$ can be written in the form given by Eq.~(\ref{gws}) with a suitable choice for $\varrho_n$. We can use the same kind of approximation to extend Eq.~(\ref{TWDp1}) for arbitrary $n$ to
\begin{equation}\label{TWDpn}
p^{(n)}(s)=\frac{1}{\tilde{f}(c_n)}s^{\varrho_n}e^{-\Gamma_n\,s^2-\beta\,v(s;{\cal A}_n)-c_n\,s},
\end{equation}
with $\varrho_n$ given by Eq.~(\ref{rhon}). In this way, for ${\cal A}=0$ we recover the case of non-interacting steps and for $n=0$ we arrive to the TWD given by Eq.~(\ref{TWDp1}). In Fig.~\ref{gr} a) are shown the results given by Eqs.~(\ref{greq}), (\ref{rhon}) and, (\ref{TWDpn}) for the case of interacting steps with $v(s;{\cal A})={\cal A}/s^2$. Fig. \ref{gr} b) shows the same results for the potential given by Eq. ~(\ref{vab}). The agreement is excellent in both cases. As expected, large values of ${\cal A}$ give better-defined peaks in $g(r)$ than in the case of non-interacting steps given by Eq.~(\ref{grff}). We also check the quality of the fit for each $p^{(n)}(s)$ with $n\ge1$, finding excellent agreement with the numerical results. These fits are not included in the text.

\subsection{Next-nearest-neighbor interactions $q=2$}
In the previous section we discussed the applicability of Eq.~(\ref{TWDp1}) for arbitrary potentials and nearest-neighbor interactions $q=1$. If we include next-nearest neighbor interactions, the effective Hamiltonian of the system takes the form
\begin{eqnarray}\label{hpeff2}
H_{\mathrm{eff}}&=&\frac{1}{\beta}\sum^{N}_{i=1}\left[K S^2_i-\mathrm{ln}\left(S^2_i\right)\right]\nonumber\\
&+&\sum^{N}_{i=1}\left(\tilde{V}(S_i;{\cal A})+\tilde{V}(S_i+S_{i+1};{\cal A})\right).
\end{eqnarray}
Then, the partition function for this system can be written as
\begin{equation}
Z(L_x)=\int^{\infty}_0 dS_1 \cdots \int^{\infty}_0 dS_N \delta(\Delta)\prod^{N}_{j=1}f(S_j)h(S_j+S_{j+1}),
\end{equation}
where $f(s)$ is given by Eq.~(\ref{f}) and
\begin{equation}
h(S)=e^{-\beta\,\tilde{V}(S;{\cal A})}.
\end{equation}

Again, following Bogomolny et al. \cite{Bogomolny}, the TWD can be written as
\begin{equation}
P(s)=\frac{L_x}{N}\left[\phi_0\left(c,\frac{s\,L_x}{N}\right)\right]^2,
\end{equation}
where $\phi_0(t,S)$ is the eigenfunction associated with the largest eigenvalue, $\lambda_0$, of the following homogeneous Fredholm integral equation
\begin{equation}\label{eqintp2}
\int^{\infty}_0 dS' K(S,S')\phi_j(t,S')=\lambda_{j}\phi_j(t,S)
\end{equation}
where the symmetric kernel, $K(S,S')$, has the form
\begin{equation}
K(S,S')=\sqrt{f(S)}e^{-\frac{t\,S}{2}}h(S+S')e^{-\frac{t\,S'}{2}}\sqrt{f(S')}.
\end{equation}
As in Eq.~(\ref{ecc}), $c$ is the solution of an algebraic equation
\begin{equation}
0=\frac{L_x}{N}+\left.\frac{1}{\lambda_{0}(t)}\frac{d\lambda_{0}(t)}{dt}\right|_{t=c}.
\end{equation}
In general, Eq.~(\ref{eqintp2}) is difficult to solve analytically. However, it can be solved numerically as a standard eigenvalue problem \cite{baker}. To clarify this point, we note that Eq.~(\ref{eqintp2}) can be written approximately as
\begin{equation}\label{eigen}
\delta \sum^{M}_{k=1} W_k K(S_l,S'_{k})\phi_j(S_k')=\lambda_{j}\phi_j(S_l).
\end{equation}
with $\delta=S'_{k+1}-S'_{k}$ the constant interval between pivotal points and $W_k$ a weighting coefficient. The original integration domain $[0,\infty)$ is approximated by $[\delta,M\,\delta]$. Eq.~(\ref{eigen}) represents a set of algebraic equations given explicitly by
\begin{equation}\label{eqintp2dis}
h\,\textbf{K}\textbf{W}\mbox{\boldmath$\phi$}_j=\lambda_{j}\mbox{\boldmath$\phi$}_j,
\end{equation}
where $\mbox{\boldmath$\phi$}^T_j=(\phi_j(h\,\delta),\phi_j(2 h\,\delta),\cdots,\phi_j(M\,\delta))$, $K(S_l,S'_k)=K(l\,\delta,k\,\delta)$ and $\textbf{W}$ is a diagonal matrix with elements $W_1, W_2,\cdots,W_M$. Naturally, the vector $\mbox{\boldmath$\phi$}_j$ gives the values of the function $\phi_j(S)$ at positions $S=l\,\delta$ with $l=1,2,\cdots,M$.

\begin{figure*}[htp]
\begin{center}
$\begin{array}{cc}
\includegraphics[scale=0.26]{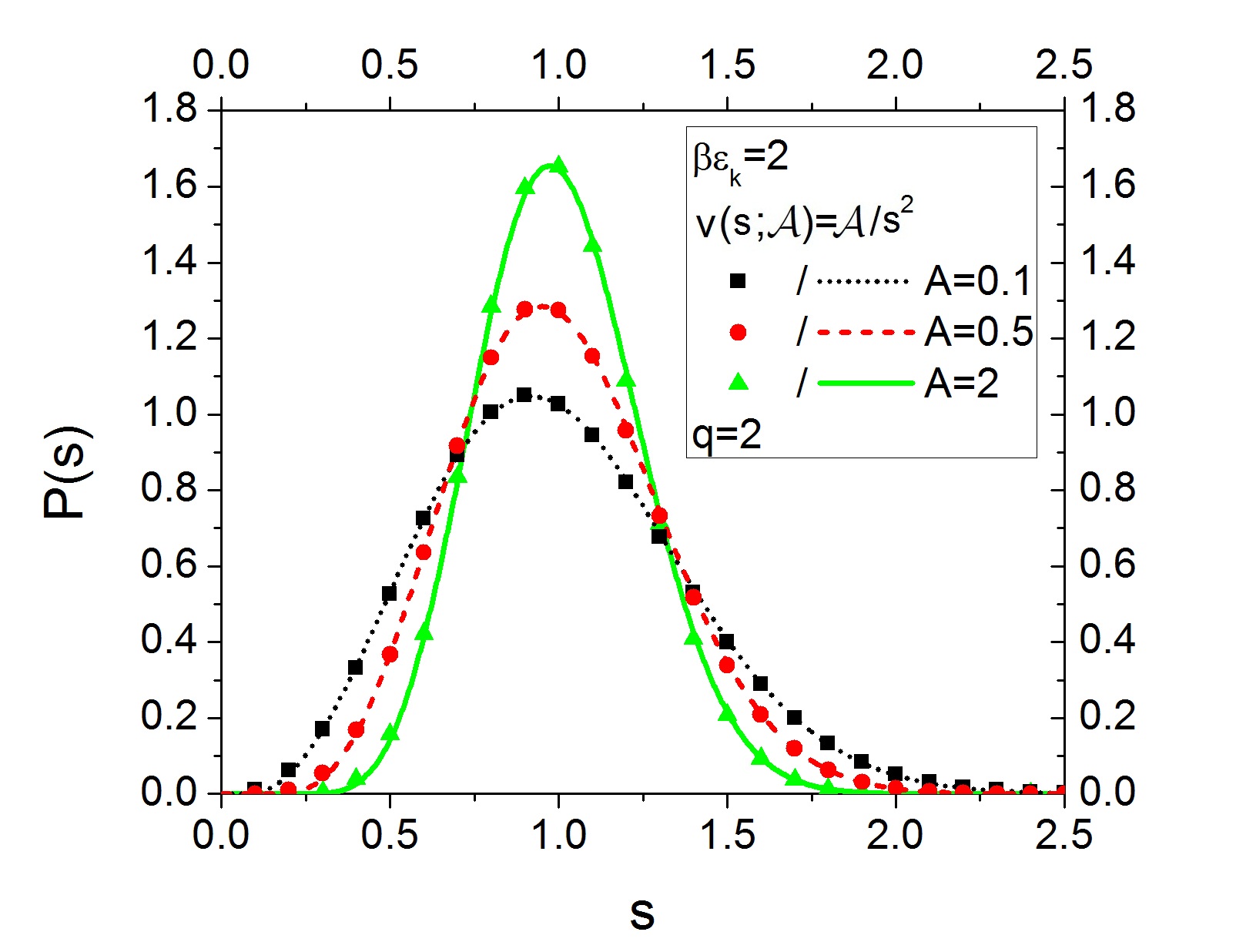} &
\includegraphics[scale=0.26]{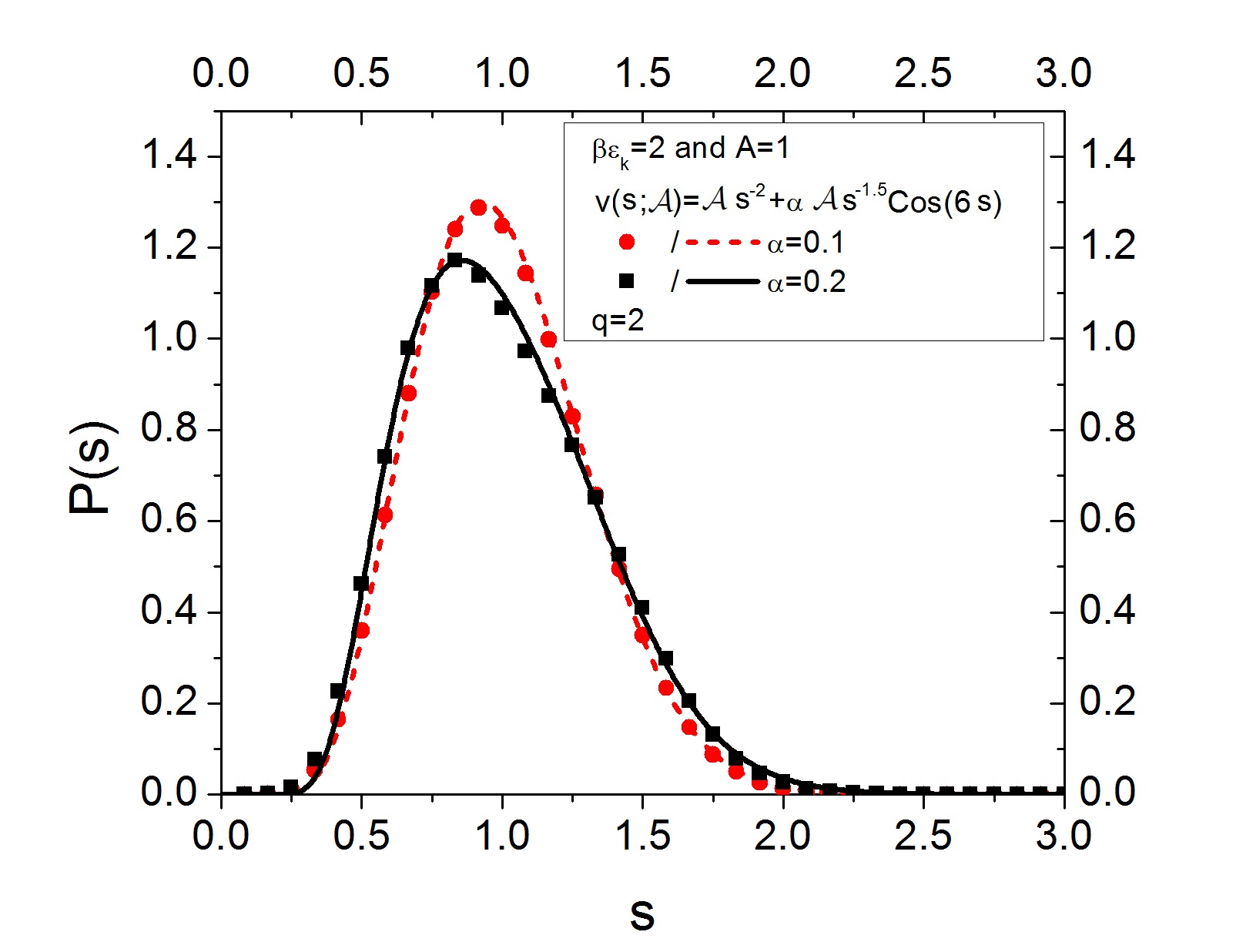}\\
(a) & (b) \\
\end{array}$
\end{center}
\caption{(Color online) a) TWD for different values of $A$ with $v(s;{\cal A})={\cal A}/s^2$ with $q=2$ and b) TWD for different values of ${\cal A}$ and $\alpha$ with $v(s;A,\alpha)=\frac{A}{s^2}+\frac{\alpha\,A\,\cos(6\,s)}{s^{3/2}}$ also with $q=2$.}
\label{pfitap2}
\end{figure*}

In order to test the quality of the solutions given by Eq.~(\ref{eqintp2dis}), we perform a comparison with numerical data from the simulation of the TSK model, displayed in Fig.~\ref{pfitap2}. As in the case of $q=1$, the agreement is excellent.

%\begin{figure}[htp]
%\begin{center}
%\includegraphics[scale=0.28]{pfitap2}\\
%\caption{(Color online) TWD for different values of $A$ with $v(A,s)=A/s^2$ and next nearest neighbor interactions $q=2$.}
%\label{pfitap2}
%\end{center}
%\end{figure}

\subsection{Arbitrary range of interactions}
\begin{figure*}[htp]
\begin{center}
$\begin{array}{cc}
\includegraphics[scale=0.26]{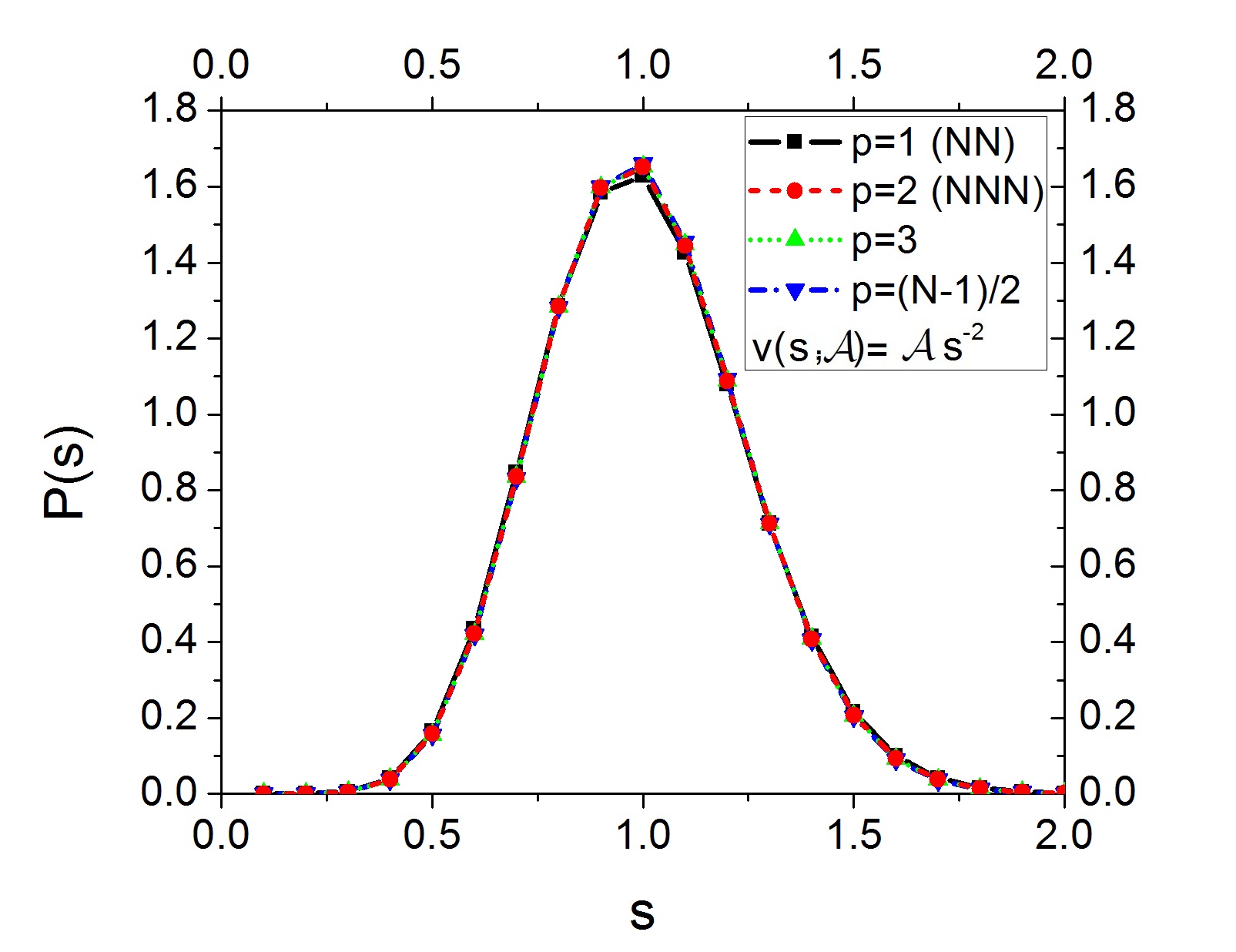} &
\includegraphics[scale=0.26]{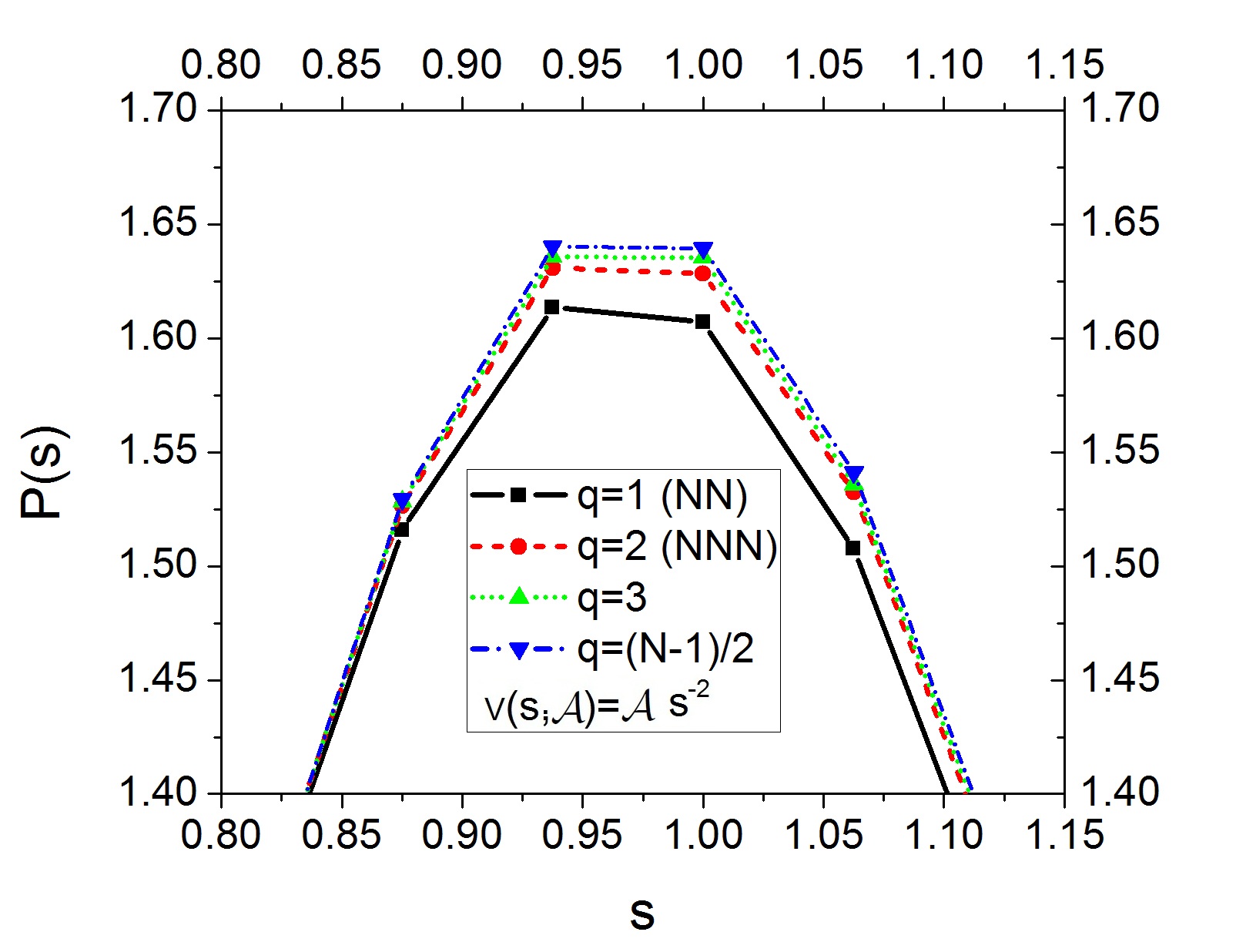}\\
(a) & (b) \\
\end{array}$
\end{center}
\caption{(Color online) a) TWD for the potential $v(s;{\cal A})={\cal A}\,s^{-2}$ for different range of interactions and b) zoom of the TWD near the maximum. For rapidly decreasing potentials we do not find significant differences in $P(s)$ with increasing $q$.}
\label{s4}
\end{figure*}

Now we consider the case where each step interacts with an arbitrary number $q$ of its neighbors. In this case $P(s)$ can be also written in terms of the eigenfunctions $\phi_j(S_1,\cdots,S_{q})$ of a complicated integral equation which involves a kernel with $2(q-1)$ variables. The resulting equation for the TWD requires the solution of $q-1$ integrals which increase in difficulty with $q$, making it hard to find a numerical solution. However, as shown in Fig.~\ref{s4} for $v(s;{\cal A})={\cal A}/s^2$, our numerical results show that the differences in the TWD for the values of $q$ considered are minor. The most significant differences are found near the maximum of the distribution. As shown in Fig. \ref{s4} b), the largest differences for the TWD are found between the cases $q=1$ and $q=2$, while the differences between the cases $q=2$ and $q=3$ or $q=3$ and full-range interactions are negligible. This is not an unexpected result because for the physically-important rapidly decreasing potentials such as $v(s;{\cal A})\propto s^{-2}$, the contribution of the interactions in the Hamiltonian are dominated by the nearest-neighbor terms. This justifies the use of our model for $q=1$ or $q=2$, even in the cases of full-range interactions. Nevertheless, we emphasize that this approximation is only valid for rapidly decreasing potentials. For potentials such as $v(s;{\cal A})=-{\cal A}\,\mathrm{ln}(s)$ the functional form $p^{(n)}(s)$ for arbitrary $n$ depends strongly on $q$ \cite{Bogomolny,gonzalezMF}.

\section{Conclusions}

The formalism presented here is quite general and can be used for a wide range of interaction potentials. We are able to describe the effect on the TWD given by electronic, elastic and entropic forces between steps. Additionally, the model allows one to describe several aspects of interacting step systems.  In particular the quantitative description of $P(s)$, $g(r)$ and $v(s;{\cal A})$ given by our formalism is remarkably good for the numerical and experimental data considered.

For potentials which decrease rapidly with $s$, in particular for the physically most important case $s^{-2}$, we found that the effect of the finite range of interactions is not significant and that the TWD can be described by taking into account just nearest-step-neighbor interactions. This means, for example, any discrepancy between findings for the TWD computed with just nearest-neighbor interactions and analytic predictions for infinite-range models (especially the GWS) must be attributed to some other source.  Nonetheless, the formalism proposed by \cite{Bogomolny} and discussed in this paper gives analytical expressions for the TWD even for arbitrary values of $q$.

Determining the interaction range $q$ from the empirical data of TWD does not seem feasible for interaction potentials which decay rapidly as $s\rightarrow \infty$ (including $s^{-2}$). The main reason is that this kind of potential affects $P(s)$ significantly just for small values of $s$, and in this limit the experimental data are too noisy to access the effects of the number of interacting neighbors. In contrast, long-range potentials such as $v(s;{\cal A})=-{\cal A}\,\mathrm{ln}(s)$ change the functional form of $P(s)$ even for large values of $s$ \cite{Bogomolny,gonzalezMF}.

\section{Acknowledgments}
This work was supported by the NSF-MRSEC at the University of Maryland, Grant No.\ DMR 05-20471, with ancillary support from the Center for Nanophysics and Advanced Materials (CNAM).

\appendix

\section{Spacing distribution functions}\label{app1}
The joint probability distribution, $P_{N}\left(x_1,\cdots,x_N;\beta\right)$, to find particles 1, 2, $\cdots$, $N$ around positions $x_1, x_2, \cdots, x_N$, respectively, is given by
\begin{equation}\label{joint}
P_{N}\left(x_1,\cdots,x_N;\beta\right)=\frac{1}{Z_N(L;\beta)}e^{-\beta\,\textbf{V}(x_1,\cdots,x_N)},
\end{equation}
where $Z_N(L;\beta)$ is the partition function of the system and $\textbf{V}(x_1,\cdots,x_N)=\sum_{i=1}^N\sum_{j=i+1}^{i+q} V(\left|x_i-x_j\right|)$ is the total interaction energy among the $N$ particles. As the interaction potential only depends on the differences between the position of the particles, the change of variables $S_i=x_{i}-x_j$ gives
\begin{equation}
P_{N}\left(S_1,\cdots\!,S_N;\beta\right)=\frac{1}{Z_N(L;\beta)}e^{-\beta\,\Omega},
\end{equation}
with
\begin{eqnarray}
\Omega &=& \sum^{N}_{m=1}\left[V(S_m)+V(S_m+S_{m+1})\right.\nonumber\\
      &&\left.+\cdots+V(S_m+\cdots+S_{m+q-1})\right].
\end{eqnarray}
The periodic boundary conditions impose $S_{N+1}=S_1$. The joint probability distribution of $n$ consecutive spacings $P_{n}\left(S_1,\cdots,S_n;\beta\right)$ can be written as
\begin{equation}
P_{n}\left(S_1,\cdots,S_n;\beta\right)=\int dS_{n+1}\cdots dS_{N}\,P_{N}\left(S_1,\cdots,S_N;\beta\right).
\end{equation}
By definition, the $n^{th}$ spacing distribution function $\hat{p}^{(n)}(S)$ can be calculated from
\begin{equation}
\hat{p}^{(n)}(S)=\int^{\infty}_{0} dS_1\cdots dS_{n+1}\,\delta\left(\eta\right)P_{n+1}\left(S_1,\cdots,S_{n+1};\beta\right),
\end{equation}
with $\eta=S-\sum^{n+1}_{i=1} S_i$. The scaled probability density is
\begin{equation}\label{pnsdef}
p^{(n)}(s)=\int^{\infty}_{0} dS_1\cdots dS_{n+1}\, \delta(\lambda)P_{n+1}\left(S_1,\cdots,S_{n+1};\beta\right).
\end{equation}
with $\lambda=\eta/\left\langle S\right\rangle$. Note that Eq.~(\ref{pnsdef}) satisfies the standard normalization conditions \cite{abul}
\begin{equation}\label{grgen}
\int^{\infty}_{0}ds\,p^{(n)}(s)=1 \quad \mathrm{and} \quad  \int^{\infty}_{0}ds\,s\,p^{(n)}(s)=n+1.
\end{equation}

\section{Configurational partition function}\label{app3}

The partition function of the system described by Eq.~(\ref{hpeff1}) is given by
\begin{equation}\label{zp1}
Z_N(L_x)=\int^{\infty}_0 dS_1 \cdots \int^{\infty}_0dS_N \delta(\Delta)\prod_{i=1}^N\,f(S_i)
\end{equation}
with $\Delta=L_x-\sum^{N}_{i=1}S_i$ and
\begin{equation}\label{f}
f(S_i)=S^2_i\,e^{-K\,S_i^2-\beta\,\tilde{V}(S_i;{\cal A})}.
\end{equation}
The Laplace transform $\tilde{Z}_N(t)=\int^{\infty}_0 dL_{x}\, e^{-t\,L_{x}}\,Z_N(L_{x})$ of Eq.~(\ref{zp1}) can be written as
\begin{equation}\label{laplaz}
\tilde{Z}_N(t)=\left(\int^{\infty}_0 dS\,e^{-t\,S}f(S)\right)^N=\left(\tilde{f}(t)\right)^N.
\end{equation}
The inverse of Eq.~(\ref{laplaz}) can be calculated by using the saddle-point approximation as shown in the Appendix \ref{app2}:
\begin{eqnarray}\label{saddle}
Z_N(L_x)&=&\frac{1}{2\,\pi\,i}\int^{\tau+i\,\infty}_{\tau-i\,\infty}dt\,e^{L_x\,t+N\,\mathrm{ln}\tilde{f}(t)}\nonumber\\
&\sim&\left(\tilde{f}(c)\,e^{\frac{L_x\,c}{N}}\right)^N,
\end{eqnarray}
where $c$ is given by the solution of the following algebraic equation
\begin{equation}\label{ecc}
0=\frac{L_x}{N}+\frac{1}{\tilde{f}(c)}\left.\frac{d\tilde{f}(t)}{dt}\right|_{t=c}.
\end{equation}
In Laplace space, the normalized TWD can be written as \cite{Bogomolny}
\begin{equation}
\tilde{P}(t)=\frac{1}{\tilde{f}(c)}\tilde{f}(c+t).
\end{equation}
Then, taking the inverse Laplace transform, we find straightforwardly
\begin{eqnarray}\label{TWDp1a}
P(s)&=&\frac{1}{\tilde{f}(c)}f(s)e^{-c\,s}\nonumber\\
&=&\frac{1}{\tilde{f}(c)}s^2e^{-\Gamma\,s^2-\beta\,v(s;{\cal A})-c\,s}
\end{eqnarray}
with $s=S\,N/L_x$ (the scaled spacing between particles), $K=\Gamma/\left\langle S\right\rangle^2$, and $\upsilon(s;{\cal A})$ is the step-step interaction potential in dimensionless form. For more information see Refs.~\cite{Bogomolny,Milan,gonzalezMF}. Until now we have not made any assumption about $\tilde{V}(S;{\cal A})$; Eq.~(\ref{TWDp1a}) applies for any potential.

\section{Saddle point approximation}\label{app2}
The integral given in Eq.~(\ref{saddle}) can be written as
\begin{equation}
Z_N(L_x)=\frac{1}{2\,\pi\,i}\int^{\tau+i\,\infty}_{\tau-i\,\infty}dt\,e^{N\,F(t)},
\end{equation}
where $F(t)=\Delta\,t+\,\mathrm{ln}\tilde{f}(t)$. Expanding $F(t)$ around $t=c$ gives
\begin{equation}
Z_N(L_x)=\frac{e^{N\,F(c)}}{2\,\pi\,i}\int^{\tau+i\,\infty}_{\tau-i\,\infty}dt\,e^{\frac{N\,F^{(2)}(c)\,(t-c)^2}{2}}G(t),
\end{equation}
with $G(t)=1+\frac{N}{4!}F^{(4)}(c)(t-c)^4+\cdots$. Additionally, $F^{(2)}(c)$ and $F^{(4)}(c)$ are the second and the fourth derivatives of $F(t)$ evaluated in $t=c$, respectively. Recall that $c$ is the solution of Eq.~(\ref{ecc}). The integral with respect to $t$ can be done choosing $\tau=c$, i.e., along the line parallel to the imaginary axis through the point $c$. This procedure gives
\begin{equation}
Z_N(L_x)\approx e^{N\,F(c)}\sqrt{\frac{1}{2\,\pi\,N\,F^{(2)}(c)}}+O\left(N^{-3/2}\right).
\end{equation}
In the thermodynamic limit $N\rightarrow \infty$ we can expect that $Z_N(L_x)\rightarrow e^{N\,F(c)}\sqrt{\frac{1}{2\,\pi\,N\,F^{(2)}(c)}}\sim \left(\tilde{f}(c)\,e^{\frac{L_x\,c}{N}}\right)^N$ where we have used the definition of $F(t)$ given previously.


\begin{thebibliography}{}
\bibitem{simu} All our numerical simulations were performed with $\beta\,\epsilon_k=2$, $N=20$ and $N_y=500$. For small values of $\beta\,\epsilon_k$ the steps are not ``smooth'' while for large values the numerical simulation is not efficient.
\end{thebibliography}
\end{document}